\documentstyle[bibnorm,epsf]{lamuphys}
\makeatletter
\let\chapter\hid@chapter
\makeatother
\begin{document}

\pagenumbering{arabic}

\title{Shot Noise Induced Charge and Potential Fluctuations of Edge States 
in Proximity of a Gate} 

\author{Markus  B\"uttiker}

\institute{D\'epartement de Physique Th\'eorique, Universit\'e de Gen\`eve,\\
CH-1211 Gen\`eve 4, Switzerland}

\titlerunning{Shot Noise Induced Charge and Potential Fluctuations ..}

\maketitle

\begin{abstract} 
We evaluate the RC-time of edge states capacitively 
coupled to a gate located away from a QPC which allows for
partial transmission of an edge channel. At long times or 
low frequencies the RC-time governs 
the relaxation of charge and current and governs the fluctuations
of the equilibrium electrostatic potential. The RC-time in mesoscopic 
structures is determined by an electrochemical capacitance which depends
on the density of states of the edge states and a charge relaxation 
resistance. 
In the non-equilibrium case, 
in the presence of transport, the shot noise leads to charge fluctuations 
in proximity of the gate which are again determined by the equilibrium 
electrochemical capacitance but with a novel resistance. The case 
of multiple edge states is discussed and the effect of a dephasing
voltage probe on these resistances is investigated. The potential fluctuations
characterized by these capacitances and resistances  
are of interest since they determine the dephasing rate in Coulomb coupled
mesoscopic conductors.  
\end{abstract}
\section{Introduction}
Dynamic fluctuations in mesoscopic conductors have attracted 
considerable attention. Most of the work has focused on 
the low frequency white
noise limit of the current fluctuations that can be measured at the
terminals of a conductor \cite{mb92}. Much less is known, if we ask about
fluctuations at higher frequencies. To be sure, there are a number of
questions which can be asked in a frequency range for which the scattering
matrix of the conductor can still be taken energy independent. All that
matters in this regime is the frequency dependence of Fermi functions which
govern the occupation of the states incident form a reservoir. Much more
interesting problems arise if we ask questions 
which directly probe the
energy dependence of the scattering matrix. 

In this work we are concerned with charge and potential fluctuations in 
Coulomb coupled systems. Such systems are of increasing interest because
one of the systems can serve as a measurement probe of the other
system \cite{field,ford}. Coulomb coupled mesoscopic systems are also of
interest in the investigation of dephasing: through the long range Coulomb
interactions the proximity of a mesoscopic conductor affects the dephasing
rate in the other conductor \cite{buks1,buks2}. The dephasing rate is
essentially determined by the fluctuations of the electrostatic potential
which leads directly to the fluctuation of the phase of a carrier. Thus a
theoretical description and experimental characterization of potential
fluctuations is essential for an understanding of such problems. Perhaps the simplest Coulomb coupled system consists of a mesoscopic
capacitor: two small plates, separated by a barrier which is too high to
permit carrier exchange, are each separately coupled to a reservoir. Such a
system permits no dc-transport, but exhibits an ac-conductance and exhibits
frequency dependent charge, potential and current fluctuations~\cite{btp,math}.
From the point of view of the scattering theory of electrical transport, it
is a simple example, in which the energy dependence of the scattering
matrix is crucial. We are not merely testing the transmission probability
of a conductor, nor the frequency dependence of the Fermi functions, but
are now asking a question that is sensitive to the charge distribution and
its dynamics. The questions we whish to address and illustrate with a simple
example in this article are of this nature. 

The dynamic behavior of a capacitor is determined by its RC-time.
At long times, the relaxation of charge and current and the electrostatic
potential is determined by this time.  
Thus it is intersting to ask: What is the RC-time of a phase-coherent
conductor? Ref.~\cite{btp,math} considered two small conductors 
each of which is connected only via a single lead to an electron 
reservoir. The two conductors interact only via the long range Coulomb 
force. Assuming that the main effect of the Coulomb interaction 
is the energy cost to charge the system, Ref.~\cite{btp} presents
an answer in terms of the geometrical capacitance and 
the energy derivatives of the scattering matrix. The resulting 
capacitance is 
called an {\it electrochemical capacitance} $C_{\mu}$,
and the resistance of the structure is called a charge 
relaxation resistance
$R_q$, to distinguish them from their 
geometrical and classical counterparts. 
Note that such a system has an 
\begin{figure}
\epsfxsize=7cm
\centerline{\epsffile{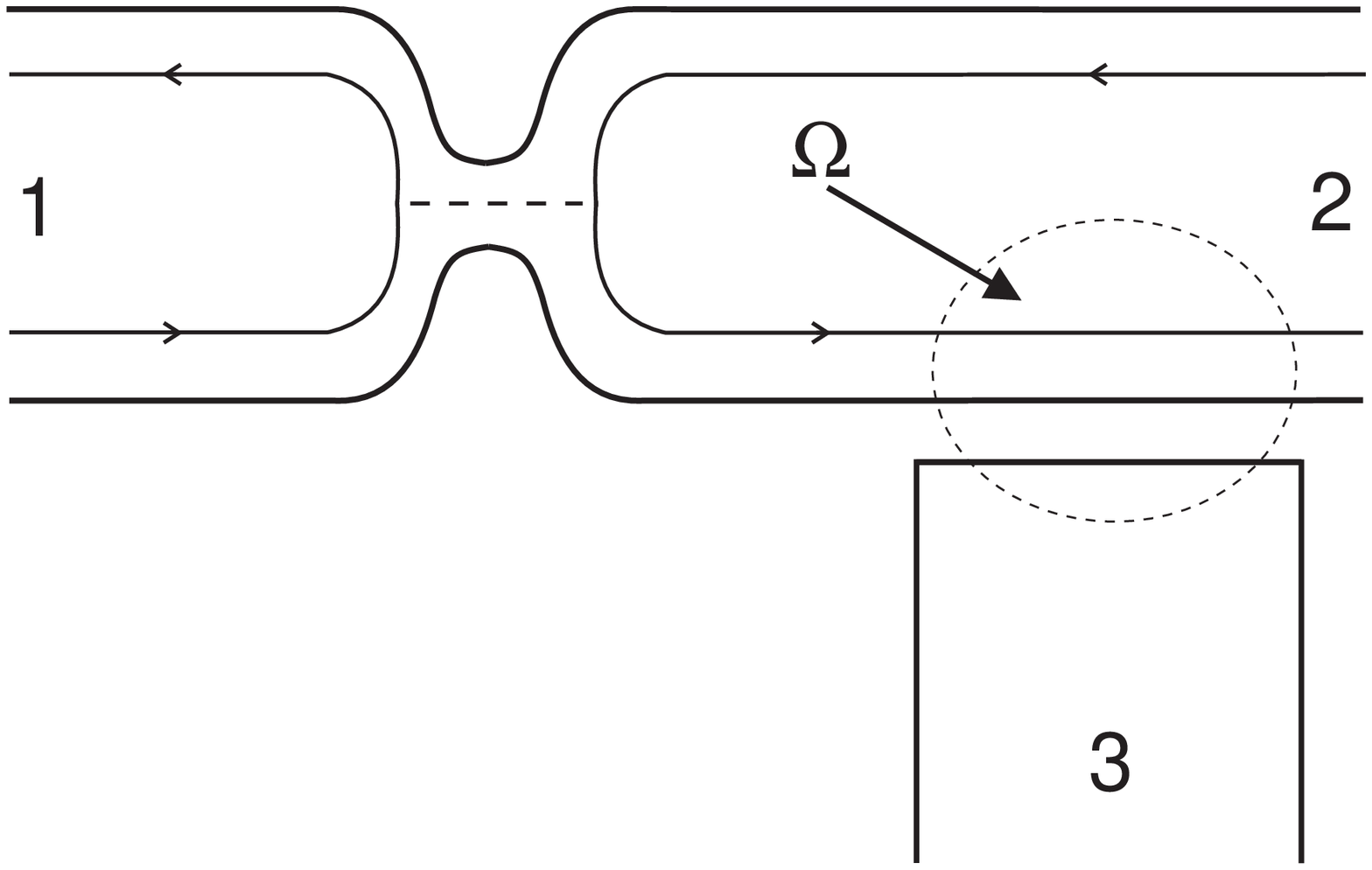}}
\caption{ \label{qpc_geometry}
Hall bar with a quantum point contact and a gate 
overlapping the edge of the conductor.
}
\end{figure}
\noindent 
infinite dc-resistance and thus the expression for the resistance $R_q$
which governs the relaxation of charge looks very different from 
the scattering matrix expressions for dc-resistances of conductors 
with non-vanishing transmission probability. 

The RC-time plays a central role also for 
mutually Coulomb coupled multiprobe conductors. 
In multiterminal structures, especially if they are ballistic, 
additional inductive-like time-scales appear \cite{christen2,bhb}. 
However, as 
soon as we consider such a conductor not in isolation, but 
coupled to another gate or conductor, the RC-time remains 
a fundamental quantity: if we keep all external potentials 
of the conductor at the same value, we are again faced with a 
purely capacitive question: To what extend can we charge 
this conductor against the other nearby conductor or gate? 

A closely related phenomenon occurs if we drive the 
conductor out of equilibrium by applying a dc voltage to it. 
Now at zero temperature the conductor exhibits shot noise \cite{leso,mb92}
which in addition to the usually investigated current 
fluctuations at the contacts 
of the conductor, generates charge fluctuations. 
These charge fluctuations depend again crucially on the 
capacitance of the mesoscopic conductor vis-a-vis other nearby 
conductors or gates. For small driving voltages,  
we find in fact that the capacitance 
is $C_{\mu}$ as in the equilibrium system. But a novel resistance 
appears \cite{plb}, which we call $R_{v}$ to indicate that it is connected 
to a non-equilibrium state obtained by applying a voltage $V$
to one of the conductors. 

The example which we treat in this work is shown in Fig. 1.
A conductor subject to a high magnetic field with a quantum point contact
(QPC) is capacitively coupled to a gate. The contacts of the conductor are
labeled $1$ and $2$ and the gate contact is labeled $3$. We assume that the
magnetic field is in a range at which the only states at the Fermi energy
which connect contacts $1$ and $2$ are edge states \cite{mb88}. 
A similar geometry
without the QPC was investigated by Chen et al. \cite{chen1}. In this work
it was shown that an oscillating voltage applied to the gate (contact $3$)
generates a current only at contact $2$ but if the magnetic field polarity
is reversed the induced current is found only at contact $1$. Since
coupling between the gate and the mesoscopic sample is purely capacitive,
this experiment verifies a prediction~\cite{mb} that capacitance
coefficients are in general not even functions of magnetic field. The
geometry with the QPC is inspired by a recent experiment of Sprinzak et
al.~\cite{buks2} which investigates the dephasing of a double quantum dot
due to the charge fluctuations generated by a current through the QPC. Here
we will consider the geometry with the gate, 
instead of the double quantum dot. 
The conductor of Fig. \ref{qpc_geometry} permits an investigation 
of the electrochemical capacitance $C_{\mu}$ and the resistances $R_q$ and 
$R_v$ of this structure. The relationship of these transport coefficients 
to the dephasing time is the subject of Ref. \cite{mbam}. We will not 
review this part of Ref. \cite{mbam} but only mention that 
related work \cite{levinson2} 
addresses this question invoking only 
the fluctuations of non-interacting electrons. 
Here we treat the fluctuations within a charge and current conserving 
self-consistent random-phase approximation (RPA) 
which represents a dynamical extension \cite{math,plb,mbam}
of Ref. \cite{levinson0}.

\section{The scattering matrix}

To be specific we consider the conductor shown 
in Fig. 1. Of interest is the current
$dI_{\alpha}(\omega)$ at contact $\alpha$ of this conductor 
if an oscillating voltage
$dV_{\beta}(\omega)$ is applied 
at contact $\beta$. Here ${\alpha}$ and ${\beta}$ label the contacts 
of the conductor and the gate and take the values $1,2,3$. Furthermore, 
we are interested in the current noise spectrum 
$S_{I_{\alpha}I_{\beta}}(\omega)$ defined as 
$2\pi S_{I_{\alpha}I_{\beta}}(\omega)\delta(\omega + \omega^{\prime}) = 1/2
\langle {\hat I}_{\alpha}(\omega){\hat I}_{\beta}(\omega^\prime) +
{\hat I}_{\beta}(\omega^\prime){\hat I}_{\alpha}(\omega) \rangle$
and the fluctuation spectrum of the electrostatic potential. 
We assume that 
the charge dynamics is relevant only 
in the region underneath the gate. Everywhere else we 
assume the charge to be screened completely. This is a strong assumption: 
In reality the QPC is made with the help
of gates (capacitors) and also exhibits its own capacitance~\cite{christen2}. 
Edge states might generate long range fields, etc. Thus the results presented 
below can only be expected to capture the main effects but can certainly 
be refined. We assume that the gate is a macroscopic 
conductor and
screens perfectly. 

The scattering matrix of the QPC alone can be described by 
$r \equiv s_{11} = s_{22} = - i {\cal R}^{1/2}$
and $t \equiv s_{21} = s_{12} = {\cal T}^{1/2}$
where ${\cal T} = 1- {\cal R}$ is the transmission probability 
through the QPC. 
Here the indices $1$ and $2$ label the reservoirs 
(see Fig. \ref{qpc_geometry}).
A carrier traversing the region underneath the gate 
acquires a phase $\phi(U)$ which depends on the 
electrostatic potential $U$ in this region. 
Since we consider only the charge pile up in this region 
all additional phases
in the scattering problem are here without relevance. The total scattering
matrix of the QPC and the traversal of the region $\Omega$ is then simply
\begin{equation} \label{sm}
{\bf s} = 
\left( \matrix{ r & t \cr 
t e^{i\phi} & r e^{i\phi} } \right).
\end{equation} 
If the polarity of the magnetic field is reversed 
the scattering matrix is given by $s_{\alpha\beta}(B) = s_{\beta\alpha}(-B) $, 
i. e. in the reversed magnetic field it is only the second column
of the scattering matrix which contains the phase $\phi(U)$. 
In what follows, the dependence of the scattering matrix on the phase $\phi$
is crucial. We emphasize that the approach presented here can be generalized 
by considering all the phases of the problem and by 
considering these phases and the amplitudes to depend 
on the entire electrostatic potential landscape \cite{math}.

\section{Density of States Matrix Elements}

To describe the charge distribution due to carriers 
in an energy interval $dE$ in our conductor, we consider the 
Fermi-field \cite{mb92}
\begin{equation}\label{fermi}
\hat{\Psi} ({\bf r},t) = \sum_{\alpha m} \int dE \psi_{\alpha m} ({\bf r},E)
\hat{a}_{\alpha m} (E) exp(-iEt/\hbar)
\end{equation}
which annihilates an electron at point ${\bf r}$ and time $t$.
The Fermi operator Eq. (\ref{fermi}) is built up 
from all scattering states $\psi_{\alpha m} ({\bf r},E)$
which have unit incident amplitude in contact $\alpha$ in channel
$m$. The operator $\hat{a}_{\alpha m} (E)$ annihilates 
an incident carrier in reservoir $\alpha$ in channel
$m$. 
The local carrier density at at point ${\bf r}$ and time $t$
is determined by $\hat{n}({\bf r},t) = \hat{\Psi}^{\dagger}({\bf r},t)
\hat{\Psi}({\bf r},t)$. 
We will investigate 
the density operator in the frequency domain, $\hat{n}({\bf r},\omega)$.
It is now very convienient and instructive to consider an expression 
for the density operator not in terms of wave functions 
but more directly in terms of the scattering matrix. 
It can be shown \cite{math}, 
that the density operator $\hat{n}({\bf r}, \omega)$, 
in the zero frequency limit,
can be written in the form 
\begin{eqnarray}\label{densop}
\hat{n}({\bf r}) = \sum_{\alpha\gamma\delta} \int dE 
\hat{a}^{\dagger}_{\gamma m} (E) n_{\gamma m\delta n} (\alpha, {\bf r}) 
\hat{a}_{\delta n} (E)
\end{eqnarray}
where the elements $n_{\gamma m\delta n}$ form a matrix of 
dimensions $M_{\gamma}*M_{\delta}$. Here $M_{\gamma}$ is the number 
of channels at the Fermi energy in contact $\gamma$.  
This matrix is given by \cite{math} 
\begin{equation} \label{elmdef}
n_{\beta \gamma} (\alpha, {\bf r} ) = 
- (1/4\pi i)
[s^{\dagger}_{\alpha\beta} (\partial s_{\alpha \gamma}/\partial eU({\bf r})) -
(\partial s^{\dagger}_{\alpha \beta}/\partial eU({\bf r})) s_{\alpha \gamma} ] .
\end{equation} 
The low frequency charge dynamics can be found if these
density of states matrix elements are known. Eq. (\ref{elmdef})
tells us that in order to find the carrier distribution and its 
fluctuations, we should introduce a small potential perturbation 
into the sample and find the scattering matrix which belongs 
to this perturbation. Clearly, such a detailed information requires
a considerable effort and even more so, if we subsequently should 
solve the Poisson equation to find the electrostatic potential 
landscape which belongs to this density distribution. To proceed we 
introduce the simplifying assumption that it is only the charge pile-up
near the gate which counts and moreover that the potential in this region 
$\Omega$ can be described with a single potential parameter $U$. 
All we need then is the density elements integrated over the region 
$\Omega$. Instead of Eq. (\ref{densop}) we want to find 
\begin{eqnarray}\label{idensop}
\hat{N}({\bf r}) &    =  & \sum_{\alpha\gamma\delta}
\int_{\Omega} d^{3}{\bf r}  \int dE 
\hat{a}^{\dagger}_{\gamma m} (E) n_{\gamma m\delta n} (\alpha,{\bf r}) 
\hat{a}_{\delta n} (E) \nonumber \\
&\equiv & 
\sum_{\alpha\gamma\delta} \int dE 
\hat{a}^{\dagger}_{\gamma m} (E) N_{\gamma m\delta n} (\alpha) 
\hat{a}_{\delta n} (E) 
\end{eqnarray}
with 
\begin{equation} \label{ielmdef}
N_{\beta \gamma} (\alpha) = 
- (1/4\pi i)
[s^{\dagger}_{\alpha\beta} (ds_{\alpha \gamma}/edU) -
(ds^{\dagger}_{\alpha \beta}/edU) s_{\alpha \gamma} ] .
\end{equation} 
Thus it is sufficient to find the variation in the scattering matrix 
for a potential that is uniform over the region of interest. 
In our example it is only 
the phase $\phi$ in Eq. (\ref{sm}) which depends on $U$. 
Thus we can evaluate the density of states elements if we know 
$d\phi/edU$. But in the WKB-limit, which is sufficient for our 
purpose,  $d\phi/edU = - d\phi/dE $. However, $d\phi/dE = 
2\pi N$ where $N$ is just the density of states of the edge state 
underneath the gate. 

We are now ready to evaluate the density of states
elements Eq. (\ref{elmdef}). 
For the specific example given by Eq. (\ref{sm}) 
we find that all elements with $\alpha =1$ vanish:
$N_{11} (1) =  
N_{21} (1) = N_{12} (1) = N_{22} (1) = 0$. 
There are no carriers incident from contact $1$ or $2$ 
which pass through region $\Omega$ and leave the conductor 
through contact $1$. The situation is different if we demand 
that the current leaves the sample through contact $2$. 
Now we find 
\begin{equation} \label{Nm}
N_{\beta\gamma}(2)  = 
\left( \matrix{ {\cal T}N & t^{\ast} r N \cr 
r^{\ast} t N & {\cal R}N} \right),
\end{equation} 
where, as already mentioned,
$N$ is the density of states 
of carriers in the edge state underneath the gate. 
For the reverse magnetic field polarity
all components of the matrix vanish except 
the elements $N_{22} (1) = {\cal T}N$ and $N_{22} (2) = {\cal R}N$. 

For the charge and its fluctuations underneath 
the gate it is not relevant through which contact carriers leave. 
The charge pile up and its fluctuations are thus governed by a matrix 
\begin{equation} \label{qdenm}
N_{\beta \gamma} = \sum_{\alpha} N_{\beta\gamma} (\alpha)
\end{equation} 
which is obtained by summing over the contact index $\alpha$ 
from the elements given by Eq. (\ref{elmdef}). 
For our example the density matrix elements for the charge are thus 
evidently given by $N_{\beta \gamma} = N_{\beta\gamma}(2)$ 
whereas for the reversed magnetic field polarity 
we have $N_{11} = {\cal T} N$, $N_{22} = {\cal R}N$ and $N_{21} = N_{21} = 0$. 

Furthermore, we will make use of the {\em injectivity} of a contact 
into the region $\Omega$ and will make use of the emissivity of
the region $\Omega$ into a contact. 
The injectivity of contact $\alpha$ is the charge injected into a region 
in response to a voltage variation at this contact,
independently through which contact the carriers leave the sample \cite{mb}. 
The injectivities of contact $1$ and $2$ are 
\begin{equation}
N_{1} =  N_{11}(1) + N_{11} (2) = {\cal T} N
\label{in1}
\end{equation}
\begin{equation}
N_{2} =  N_{22}(1) + N_{22} (2) = {\cal R} N
\label{in2}
\end{equation}
Note that the sum of the injectivities of both contacts is just the density
of states $N$ underneath the gate. 
The {\em emissivity} of region $\Omega$ is the 
portion of the density of states of carriers 
which leave the conductor through contact $\alpha$ 
irrespectively from which contact they entered
the conductor \cite{mb}. We find emissivities
\begin{equation}
N^{(1)} =  N_{11}(1) + N_{22} (1) = 0,  
\label{e1}
\end{equation}
\begin{equation}
N^{({2})} =  N_{11}(2) + N_{22} (2) = N . 
\label{e2}
\end{equation} 
Any charge accumulation or depletion is only felt in contact $2$.
The injectivities and emissivities in the magnetic field 
$B$ are related by reciprocity to the 
emissivities and injectivities in the reversed magnetic field, 
$N_{\alpha}(B) = N^{(\alpha)}(-B)$ and 
$N^{(\alpha)}(B) = N_{\alpha}(-B)$.
In contrast, the density of states
$N$ is an even function of magnetic field. 

\section{The Poisson Equation: The effective interaction}

Thus far we have only considered bare charges. The true charge, 
however, is determined by the long range Coulomb interaction. 
First we consider the screening of the average charges and in a second 
step we consider the screening of charge fluctuations. We describe the 
long range Coulomb interaction between the charge on the edge state 
and on the gate with the help of a geometrical capacitance $C$. 
The charge on the edge state beneath the gate is determined by the voltage 
difference between the edge state and the gate $dQ = C(dU - dV_{g})$,
where $dU$ and $dV_{g}$ are deviations 
from an 
equilibrium reference state.
On the other hand the charge beneath the gate can also be expressed 
in terms of the injected charges  $e^{2} N_{1} dV_{1}$  in response 
to a voltage variation at contact $1$ and $e^{2} N_{2} dV_{2}$
in response to a voltage variation at contact $2$. Furthermore, 
the injected charge leads to a response in the internal potential 
$dU$ which in turn generates a screening charge $-e^{2}NdU$
proportional to the density of states. 
Thus the Poisson equation for the charge underneath the gate is 
\begin{equation}\label{poisson1}
dQ = C (dU - dV_{g}) = e^{2} N_{1} dV_{1} + e^{2} N_{2} dV_{2} 
- e^{2}N dU
\end{equation}
and the charge on the gate is given by $- dQ = C (dV_{g} - dU)$. 
Solving Eq. (\ref{poisson1}) for $dU$ gives 
\begin{equation}\label{u1}
dU = G_{eff} (C dV_{g} + e^{2} N_{1} dV_{1} + e^{2} N_{2} dV_{2}), 
\end{equation}
where $G_{eff} = (C +e^{2}N)^{-1}$ is an effective (RPA) interaction 
which gives the potential underneath the gate in response to an increment 
in the charge. 

\section{Admittance}

Consider now the low-frequency conductance:
To leading order in the frequency $\omega$ we write 
\begin{equation} \label{admit}
G_{\alpha\beta}(\omega) = G_{\alpha\beta} (0) - i \omega E_{\alpha\beta}
+ \omega^{2}  K_{\alpha\beta} + O(\omega^{3}) .
\end{equation}
Here  $G_{\alpha\beta} (0)$ is the dc-conductance matrix,  
$E_{\alpha\beta}$ is the {\em emittance} matrix, and $K_{\alpha\beta}$
is a second order dissipative contribution to the frequency dependent 
admittance. 
The zero-frequency dc-conductance matrix has only four non-vanishing
elements which are given by 
$G \equiv G_{11} = G_{22} = - G_{12} = - G_{21} = (e^{2}/h) {\cal T}$. 
Ref.~\cite{mb} showed that the emittance matrix ${\bf E}$ is given by 
\begin{equation} \label{Em}
E_{\alpha\beta} = e^{2} N_{\beta\beta}(\alpha) - e^{2} N^{(\alpha)}G_{eff}N_{\beta}
\end{equation}
As it is written, Eq. (\ref{Em}) applies only to the elements 
where $\alpha$ and $\beta$ take the values $1$ or $2$. 
The remaining elements can be obtained from current conservation 
(which demands that the elements of each row 
and column of this matrix add up to zero) or can be obtained directly
by using a more general formula \cite{christen2}. For our example we find an emittance matrix, 
\begin{equation} \label{em}
{\bf E} = C_{\mu}
\left( \matrix{  0 &   0 &  0 \cr 
                 {\cal T} &   {\cal R} & -1 \cr  
               - {\cal T} & - {\cal R} &  1} \right),
\end{equation}
with an electrochemical capacitance of the conductor vis-a-vis the gate 
given by $C_{\mu} = C e^{2}N/(C+e^{2}N)$. 
Eq. (\ref{em}) determines the displacement currents 
in response to an oscillating voltage at one of the contacts. 
There is no displacement current at contact $1$
(the elements of the first row vanish) which is consequence of our 
assumption that charge pile up occurs only underneath the gate. 
The emittance matrix in the magnetic field $B$ and in the magnetic 
field $-B$ are related by reciprocity, 
$E_{\alpha\beta}(B) = E_{\beta\alpha}(-B)$. 
For the reverse polarity, a voltage oscillation at contact $1$ 
generates no displacement currents (the elements of the first column 
vanish). 

The emittance matrix element $E_{21}$ is positive and thus 
has the sign not of a capacitive but of an inductive 
response. The elements of row $3$ and column $3$ are a consequence of 
purely capacitive coupling and have the sign associated with the 
elements of a capacitance matrix. Thus these elements 
represent the capacitance matrix elements which can be measured 
in an ac-experiment. Note that the capacitances $E_{31} \equiv C_{31}$
and $E_{32} \equiv C_{32}$ depend not only on the density of states 
and geometrical capacitances but also on transmission and reflection
probabilities. Measurement of these capacitances provides
thus a direct confirmation of the concept of {\em partial} density 
of states~\cite{mb,christen2}. Furthermore, we see that for instance $C_{31} (B)  \equiv 
E_{31} = {\cal T} C_{\mu}$ but $C_{31} (-B) = 0$. A similarly striking variation 
of the capacitance coefficients was observed in the experiment of Chen et
al.~\cite{chen1} in the integer quantum Hall effect and in 
Refs.~\cite{johnson} in the fractional quantum Hall effect. 

\section{Bare charge fluctuations}

Let us now turn to the charge fluctuations. With the help of 
the charge density matrix the low frequency limit of the 
bare charge fluctuations can be obtained~\cite{btp,plb,mbam}. It is given by 
\begin{equation}
    S_{NN}(\omega) = h 
        \sum_{\delta\gamma} \int dE\, F_{\gamma\delta}(E,\omega) 
      Tr[  N_{\gamma\delta}(E,E+\hbar\omega)
        N^{\dagger}_{\gamma\delta} (E,E+\hbar\omega) ]
\label{qfluct1}
\end{equation}
where the elements of $N_{\gamma\delta}$ are in the zero-frequency limit 
of interest here given by Eq. (\ref{qdenm})  
and $F_{\gamma\delta} = f_{\gamma}(E)(1-f_{\delta}(E+\hbar \omega))
+ f_{\delta}(E+\hbar \omega)(1- f_{\gamma}(E))$
is a combination of Fermi functions. 
Using only the zero-frequency 
limit of the elements of the charge operator determined above gives, 
\begin{eqnarray}\label{qfluct2}
    S_{NN}(\omega) =  h N^{2} &[ & {\cal T}^{2} \int dE\ F_{11}(E,\omega) 
                            + {\cal T}{\cal R} \int dE\ F_{12}(E,\omega)\nonumber\\
                         &+ & {\cal T}{\cal R} \int dE\ F_{21}(E,\omega)
                            + {\cal R}^{2} \int dE\ F_{22}(E,\omega)] .
\end{eqnarray}
At equilibrium all the Fermi functions are identical and we obtain 
$ S_{NN}(\omega) = h N^{2} \int dE\ F(E,\omega)$ 
which in the zero-frequency limit is 
\begin{eqnarray}
    S_{NN}(\omega) = h N^{2} kT 
\label{qfluct3a}
\end{eqnarray}
and at zero-temperature to leading order in frequency is, 
\begin{eqnarray}
    S_{NN}(\omega) = h N^{2} \hbar \omega .
\label{qfluct3b}
\end{eqnarray}
In the zero-temperature, zero-frequency limit,
in the presence of a current through the sample, we find 
for the charge fluctuations associated with shot noise 
\begin{eqnarray}
    S_{NN}(\omega) = h N^{2} {\cal T}{\cal R}  e|V| .
\label{qfluct4}
\end{eqnarray}
However, the bare charge fluctuations are not by themselves 
physically relevant.

\section{Fluctuations of the true charge}

To find the fluctuations of the true charge 
we now write the Poisson equation for the fluctuating charges. 
All contact potentials are at their equilibrium value, 
$dV_{1} = dV_{2} = dV_{g} = 0$. 
The fluctuations of the bare charge now generate fluctuations
in the electrostatic potential. Thus the electrostatic potential 
has also to be represented by an operator ${\hat U}$. Furthermore, the potential 
fluctuations are also screened. As in the case of the average charges we take 
the screening to be proportional to the density of states $N$ but replace 
the c-number $U$ by its operator expression ${\hat U}$. 
The equation for the fluctuations of the true charge is thus 
\begin{equation} \label{poisson2}
d {\hat Q} = C d {\hat U}  = e\hat{\cal N} - e^{2}N {\hat U}
\end{equation}
whereas the fluctuation of the charge on the gate is simply
$ - d \hat Q = - C d \hat U$.
 
Thus $d\hat Q$ is the charge operator 
which determines the {\em dipole} which forms 
between the charge on the edge state and the charge on the gate. 
Solving Eq. (\ref{poisson2})
for the potential operator ${\hat U}$ and using this result 
to find the fluctuations of the charge $d{\hat Q}$ gives 
\begin{equation}\label{qfluct1b}
    S_{QQ}(\omega) = e^{2} C^{2}G^{2}_{eff}  S_{NN}(\omega) 
                   =  2C^{2}_{\mu} (1/2e^{2}) 
                   (S_{NN}(\omega)/N^{2}) .
\end{equation}
We now discuss three limits of this result. 

\section{Equilibrium and non-equilibrium charge relaxation resistance} 

At equilibrium, in the zero-frequency limit, the charge fluctuation 
spectrum can be written with the help of the equilibrium charge relaxation
resistance \cite{btp,math,plb} $R_{q}$,  
\begin{eqnarray}
    S_{QQ}(\omega) = 2C^{2}_{\mu} R_q kT.  
\label{qflucteq}
\end{eqnarray}
For our specific example\cite{mbam}, we find using Eqs. (\ref{qfluct3a}) and 
(\ref{qfluct1b}),
\begin{equation} \label{rq}
R_q = h/2e^{2} .
\end{equation} 
The charge relaxation resistance is universal 
and equal to {\em half} a resistance quantum as expected for a single
edge state~\cite{christen2}. 
At equilibrium the fluctuation spectrum is via the fluctuation 
dissipation theorem directly related to the dissipative part 
of the admittance. We could also have directly evaluated 
the element $K_{33}$ of Eq. (\ref{admit}) 
to find $K_{33} = C_{\mu}^{2} R_{q}$.
Second at equilibrium, but for frequencies which are large 
compared to the thermal energy, but small compared to any intrinsic excitation
frequencies, we find that zero-point fluctuations give rise 
to a noise power spectral density
\begin{eqnarray}
    S_{QQ}(\omega) = 2C^{2}_{\mu} R_q \hbar \omega  
\label{qfluctz}
\end{eqnarray}
which is determined by the charge relaxation resistance Eq. (\ref{rq}). 
Third, in the presence of transport, we find in the zero-frequency,
zero-temperature limit, 
a charge fluctuation spectrum \cite{plb}, 
\begin{eqnarray}
    S_{QQ}(\omega) = 2C^{2}_{\mu} R_{v} e|V|, 
\label{qfluctv}
\end{eqnarray}
where $|V|$ is the voltage applied between the two contacts of the sample
and a non-equilibrium charge relaxation resistance which for our 
example is given by \cite{mbam}
\begin{eqnarray}
      R_{v} = (h/e^{2}) {\cal T}{\cal R}.
\label{rqv}
\end{eqnarray}
It is maximal for a semi-transparent QPC, ${\cal T} = {\cal R} =1/2$.

The current at the gate due to the charge fluctuations 
is $dI_{g} = -i\omega dQ(\omega)$ and thus its fluctuation 
spectrum is given by  
$S_{I_{g}I_{g}}(\omega) = \omega^{2} S_{QQ}$. The potential 
fluctuations are related to 
the charge fluctuations by $d{\hat U} =d{\hat Q}/C$ and thus 
the spectral density of the potential 
fluctuations is $S_{UU}(\omega) = C^{-2} S_{QQ}$. 
Thus the charge relaxation resistance determines, together with 
the electrochemical and geometrical capacitance, 
the fluctuations of the charge,
the potential and the current induced into the gate. Since dephasing 
rates can be linked to the low frequency limit of the potential 
fluctuations~\cite{aak} the 
resistances $R_{q}$ and $R_{v}$ also determine the 
dephasing rate in Coulomb coupled mesoscopic conductors~\cite{mbam}.

\section{Several edge states}

Let us next consider the case, where there are several edge states.
A QPC in a high magnetic field permits perfect transmission of the outer
edge states (belonging to the lower Landau levels) and it is only the
innermost edge state which is partially transmitted or reflected at the
QPC. 
Let us just consider two edge states: the outer edge state labeled $1$ is
perfectly transmitted $T_{1} =1$, whereas the inner edge state labeled $2$
has a transmission probability $T_{2} \equiv T$ which might take any value
between zero and one. The outer edge state, with transmission probability
$1$ is entirely noiseless as far as the shot noise in the total current is
concerned \cite{mb92}.
One might thus be tempted to think that such a perfectly transmitted edge
state plays no role at all. That however is not the case. Our result
involves screening in an essential manner and the charge fluctuations in one
of the edge states can now be screened by charge accumulation or depletion in
the other edge state. The screening properties depend on the electrostatic
interaction between the two edge states. Thus the answer we obtain depends
on the detailed electrostatic assumptions which we invoke to treat this
problem. Here, to provide a simple discussion, we assume that the two
edge states are so close, that they can be described with a common
electrostatic potential $U$. If we denote the density of states of the edge
states $1$ and $2$ in the region $\Omega$ of interest by $N_{1}$ and
$N_{2}$ a detailed consideration, repeating the procedure given above for
one edge state only, leads to an equilibrium 
charge relaxation resistance \cite{mbam}
\begin{equation} \label{rq2}
R_q = \frac{h}{2e^{2}} \frac{N_{1}^{2}+N_{2}^{2}}{(N_{1}+N_{2})^{2}}
\end{equation} 
Note that in contrast to the single edge state, now $R_q$ depends
explicitly on the densities of states. We can expect that the density of
states $N_{2}$ of the inner edge state $2$ is typically larger than the
density of states of the outer edge state since the potential for the inner
edge state is much shallower. In this case $N_{2} >> N_{1}$ and $R_q$ for
the two edge states will in fact be the same as for one edge state only. In
contrast, for samples with a sharp edge, we can expect that both density of
states are comparable, and thus $R_{q}$ for two edge states will be nearly a
factor $2$ smaller than the $R_{q}$ of a single edge state only. 

Similarly, if we investigate $R_v$ for two edge states, 
we find \cite{mbam}
\begin{equation} \label{rq22}
R_{v}= \frac{h}{2e^{2}} \frac{N_{2}^{2}}{(N_{1}+N_{2})^{2}} TR .
\end{equation}
Again the density of states of the two edge states appear now explicitly.
The density of states of the outer edge state appears only in the
denominator since it plays a role only in screening but it is not a primary
source of charge fluctuations. In the limit $N_{2} >>N_{1}$ of a shallow
edge the outermost edge state is unimportant, whereas for a steep edge if
both density of states are comparable, $R_v$ is reduced by a factor $4$
compared to the case of a single edge state only.

Form the above results it is obvious how the formulas must be written if
there is one edge state which is partially reflected or transmitted and
many edge states which are perfectly transmitted.

\section{Phase Randomization} 

\begin{figure}
\epsfxsize=7cm
\centerline{\epsffile{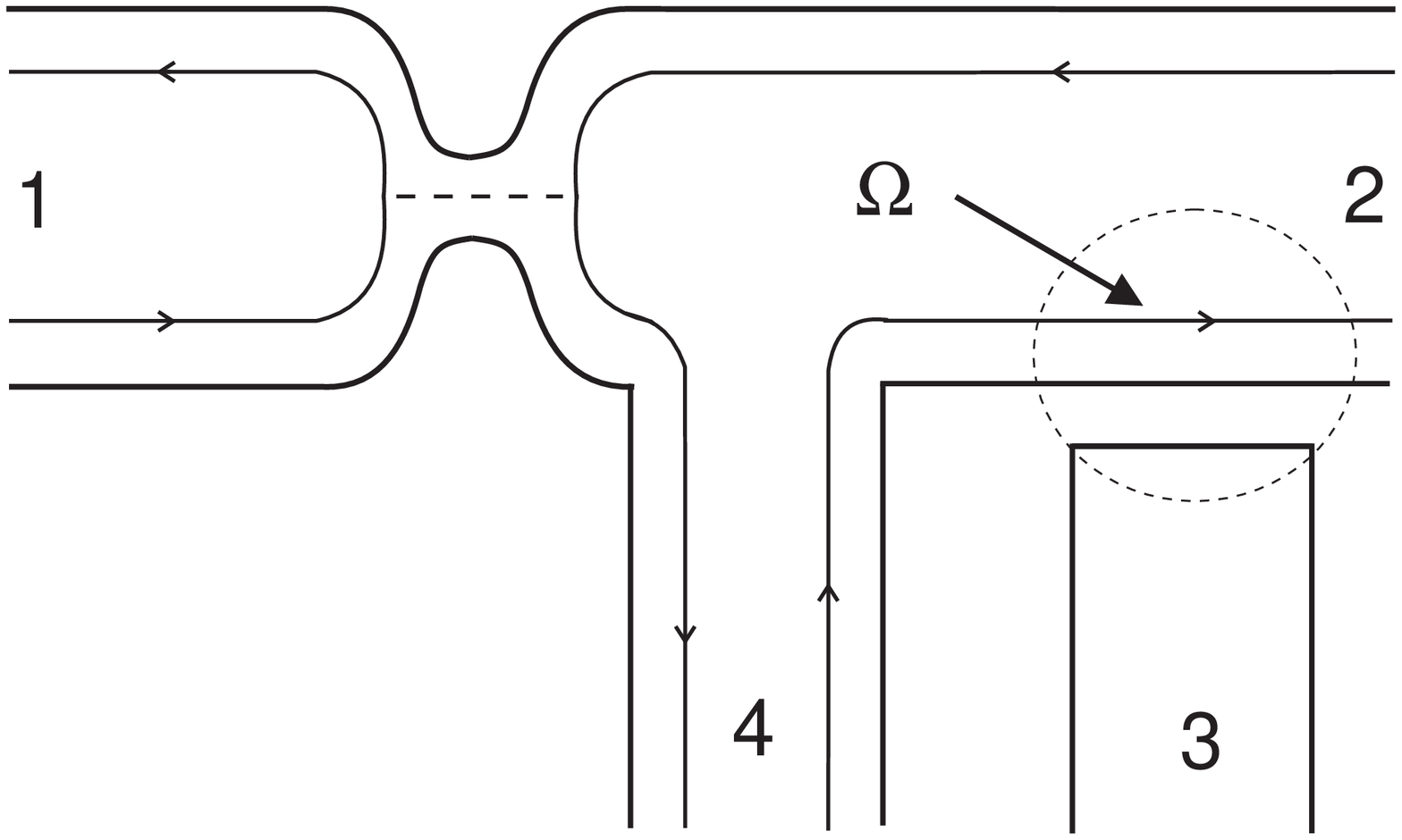}}
\caption{ \label{volt-geometry}
Hall bar with a quantum point contact and a gate 
overlapping the edge of the conductor. A voltage contact 
between the QPC and the gate serves to destroy 
quantum coherent electron motion along the edge states.
}
\end{figure}

Is the result given above sensitive to phase? 
Experimentally this question is investigated 
by Sprinzak et al. \cite{buks2}. 
Our result for one channel, Eq. (\ref{rqv}), contains only transmission 
probabilities.  To investigate this question, 
we consider, like the experiment, an additional contact
between the QPC and the region $\Omega$ as shown in Fig. \ref{volt-geometry}.
The contact will be considered as a voltage probe. An ideal voltage probe
exhibits infinite impedance at all frequencies. 
Consequently, the net current at the
voltage probe vanishes at every instant of time. Thus the voltage of the
probe becomes a fluctuating quantity. Despite the fact that the total
current vanishes, carriers leave the sample through this contact, and are
replaced by carriers which enter from the reservoir. Carriers leaving into
the reservoir and carriers rentering the conductor from the reservoir have
no phase relationship and consequently a voltage probe 
acts as a dephasor \cite{mb88}. 

The voltage probe changes the conductor: if we include the gate we
now deal with a four probe conductor. We keep for the gate the label $3$
and designate the voltage probe as contact $4$. Since the potential 
is a function of time, we must also know the dynamic conductance of the system.
To begin we consider the general relation between currents 
and voltages of our four-

\noindent 
terminal conductor. This relation 
takes the form of a Langevin equation which includes 
the fluctuating currents at the terminals as noise sources \cite{mb92}
\begin{equation} \label{langevin}
dI_{\alpha}(\omega) = \sum_{\beta} G_{\alpha\beta}(\omega) dV_{\beta}(\omega)
+ \delta I_{\alpha} (\omega) .
\end{equation}
Here $G_{\alpha\beta}(\omega)$ is the self-consistent 
dynamic conductance and $\delta I_{\alpha} (\omega)$ are the 
(self-consistent) frequency-dependent current fluctuations 
at the contacts of the conductor. 
Since the current spectrum at the 
gate is second order in frequency, it is sufficient to 
calculate the current amplitudes to first order in frequency. 
We thus need $G_{\alpha\beta}(\omega)$ only to first order 
in frequency and write $G_{\alpha\beta}(\omega)
= G_{\alpha\beta}(0) -i\omega E_{\alpha\beta}+ O(\omega^{2})$.
Here $G_{\alpha\beta}(0)$ is the dc-conductance which 
for $\nu-1$ perfectly transmitted channels and one
partially transmitted channel at the QPC is 
given by $G_{11} = -G_{12} = -G_{41} = (e^{2}/h) (\nu -1 +T)$, 
$G_{22} = G_{44} = - G_{24} = - (e^{2}/h)\nu$ and $G_{42} = -(e^{2}/h)R$. 
All other elements vanish. Repeating the calculation which led to Eq. (\ref{Em})
for the conductor of Fig. (\ref{volt-geometry}), we find
$E_{23} = -E_{24} = E_{33} = -E_{34} = -i\omega C_{\mu}$
with $C_{\mu}$ as given in Eq. (\ref{Em}).
Inserting these results into Eq. (\ref{langevin}) and holding 
all potentials, except $dV_{4}$ at their equilibrium value
gives for $I_{3}$ and $I_{4}$,  
\begin{eqnarray} \label{I4}
I_{3} & = & -i\omega C_{\mu} dV_{4} \nonumber \\
I_{4} & = & \frac{e^{2}}{h} \nu dV_{4} + \delta I_{4}
\end{eqnarray}
The noise spectrum at the voltage probe at low frequencies 
is just the spectrum of the noise of a QPC 
$S^{0}_{I_{4}I_{4}}(\omega) = 2 \frac{e^{2}}{h} TR e|V| $
where we have added an upper index $0$ to indicate that it is 
the spectrum for zero external impedance. Note that there 
is no noise source to order $\omega$ in the total current 
for $I_{3}$. (The lowest order in frequency which is dissipative 
is proportional to $\omega^{2}$). For an ideal (infinite impedance)
voltage probe we have $I_{4} = 0$ and consequently 
\begin{equation} \label{mu4}
dV_{4} (\omega) = - \frac{h}{e^{2}\nu} \delta I_{4}(\omega)
\end{equation}
Inserting this result in the equation for $I_{3}$ we find 
$S_{I_{3}I_{3}}(\omega) = 
\omega^{2} C_{\mu}^{2} S^{0}_{I_{4}I_{4}} (\omega) /\nu^{2}$. 
Using the shot noise power spectrum for $S^{0}_{I_{4}I_{4}} (\omega)$
gives for the spectrum at the gate 
$S_{I_{3}I_{3}} (\omega) = $
$2 \omega^{2} C_{\mu}^{2} R_{v} e|V|$ with \cite{mbam}
\begin{equation} \label{rqdep}
R_{v} = \frac{h}{e^{2}} \frac{1}{\nu^{2}} TR 
\end{equation}
Eq. (\ref{rqdep}) makes now an
interesting prediction. For one edge state only, the
dephasing voltage probe has no effect. The fluctuations observed at the gate
remain unchanged. If there are several edge states, the voltage probe does
have an effect since the voltage probe re-injects an equal current
into all edge states. 
The difference between Eq. (\ref{rqdep}) and
Eq.(\ref{rq2}) is, however quite subtle. $R_{v}$ as given above is simply
inversely proportional to the square of the number of edge states. Without
the voltage probe we have seen that 
$R_{v}$ varies between $\frac{1}{e^{2}} TR $ for a steep edge and
$\frac{1}{4e^{2}} TR $ for a shallow edge. Thus for a steep edge introducing 
a voltage probe has a considerable effect, whereas for a shallow edge
introducing a voltage probe has no effect at all. 

Apparently, in the experiment \cite{buks2} the voltage probe is 
not ideal. 
Instead of an infinite impedance it might, at the relevant frequency,
exhibit a finite
impedance $Z_{ext}(\omega)$. 
We assume that the external impedance 
arises from a macroscopic
circuit and its noise is voltage independent. In the presence of a
finite impedance the current Eq. (\ref{I4}) 
can also be expressed as $I_{4} = -Z^{-1}_{ext} (\omega) \delta V_{4}$.
Consequently, instead of Eq. (\ref{mu4}) we find 
\begin{equation} \label{muext}
\delta V_{4} = - \frac{Z_{ext}}{1+G_{0}Z_{ext}} \delta I_{4}
\end{equation}
where we have introduced the abbreviation $G_{0} = \nu e^{2}/h$.
Repeating the considerations given above, we find for the resistance $R_{v}$

\begin{equation} \label{rqext}
R_{v} = \frac{e^{2}}{h}
\frac{|Z_{ext}|^{2}}{|1+G_{0}Z_{ext}|^{2}} TR 
\end{equation}

This consideration shows that a finite external impedance reduces
the current fluctuations induced into the gate. Clearly this is simply a
consequence of the fact that for a finite external impedance part of the
current is "lost" at the voltage probe. This effect becomes significant
when $Z_{ext}(\omega)$ at the frequency of interest becomes 
smaller than $G_{0}^{-1}$.

\section{Discussion}

In this work, we have illustrated the calculation of charge and potential
fluctuations for a simple problem: A Hall conductor with a QPC has on
its side a gate which couples capacitively to the edge states. We have asked:
What is the current induced into this gate due to the shot noise generated
a the QPC. The simplifying assumption we have made is
that the conductor remains charge neutral everywhere
except near the gate where a charge pile-up limited by the Coulomb
interaction between gate and 
edge is permitted. This allows a solution in terms of one fluctuating 
potential only. 

Independent of the detailed discussion it is clear that the 
non-equilibrium resistance $R_v$ reflects the shot noise. The theoretical 
question concerns only the factor of proportionality. 
If we measure $R_v$ in units of $R_0 = h/2e^{2} \cal{T} \cal{R}$, 
we find that for one edge state 
$R_v/R_0$ is universal, wheras in the presence of a number of edge 
states it is not-universal, except if an ideal voltage probe 
completely equilibrates different channels, in which case we find 
$R_v/R_0 = 1/\nu^{2}$, where $\nu$ is the number of edge states. 
In Ref. \cite{buks2} it is argued that the dephasing rate (which 
is proportional to $R_{v}$) should be {\it periodic} in a phase
with period $\pi$ even for a single edge state. In contrast, in our
our result \cite{mbam}, 
Eq. (\ref{rqv}), such a periodic factor does not appear. We conclude by 
mentioning that the approach out-lined here can be generalized to hybrid
normal-superconducting systems \cite{amtgmb}.

\section{Acknowledgments}

This work was supported by the Swiss National Science Foundation
and by the TMR network Dynamics of Nanostructures.\\

%

%
%

\end{document}